\documentclass[fleqn,usenatbib]{mnras}

\usepackage{newtxtext,newtxmath}

\usepackage[T1]{fontenc}

\DeclareRobustCommand{\VAN}[3]{#2}
\let\VANthebibliography\thebibliography
\def\thebibliography{\DeclareRobustCommand{\VAN}[3]{##3}\VANthebibliography}

\usepackage{graphicx}	
\usepackage{amsmath}	
\usepackage{pifont}
\usepackage{mathtools}
\usepackage{placeins}
\usepackage{float}
\usepackage{bm}
    
\usepackage{nccmath}

\usepackage{cleveref}
\crefname{figure}{Fig.}{Figs.}
\crefname{table}{Table}{Tables}

\title[Non-ideal magnetohydrodynamics on a moving mesh]{Non-ideal magnetohydrodynamics on a moving mesh II: Hall effect}

\author[O. Zier, A. C. Mayer and V. Springel]{%
Oliver Zier$^{1}$\thanks{E-mail: ozier@mpa-garching.mpg.de}, 
 Alexander C. Mayer$^{1}$ and Volker Springel$^{1}$
\vspace*{0.1cm}\\%
$^{1}$Max-Planck-Institut für Astrophysik, Karl-Schwarzschild-Straße 1, 85741 Garching, Germany\\
}

\date{Accepted XXX. Received YYY; in original form ZZZ}

\pubyear{2023}

\begin{document}
 \label{firstpage}
\pagerange{\pageref{firstpage}--\pageref{lastpage}}
\maketitle

\begin{abstract}
In this work we extend the non-ideal magnetohydrodynamics (MHD) solver in the moving mesh code {\small AREPO} to include the Hall effect. The core of our algorithm is based on an estimation of the magnetic field gradients by a least-square reconstruction on the unstructured mesh, which we also used in the companion paper for Ohmic and ambipolar diffusion. In an extensive study of simulations of a magnetic shock, we show that without additional magnetic diffusion our algorithm for the Hall effect becomes unstable at high resolution. We can however stabilise it by artificially increasing the Ohmic resistivity, $\eta_{\rm OR}$, so that it satisfies the condition $\eta_{\rm OR} \geq \eta_{\rm H} /5$, where  $\eta_{\rm H}$ is the Hall diffusion coefficient. Adopting this solution we find second order convergence for the C-shock and are also able to accurately reproduce the dispersion relation of the whistler waves.  As a first application of the new scheme, we simulate the collapse of a magnetised cloud with constant Hall parameter $\eta_{\rm H}$ and show that, depending on the sign of $\eta_{\rm H}$, the magnetic braking can either be weakened or strengthened by the Hall effect. The quasi-Lagrangian nature of the moving mesh method used here automatically increases the resolution in the forming core, making it well suited for more realistic studies with non-constant magnetic diffusivities in the future.
\end{abstract}


\section{Introduction}
Magnetic fields play an important role at different scales in the Universe. They can interact with a conducting fluid, decisively influence the evolution of the fluid \citep{ferriere2001interstellar, cox2005three} and at the same time amplify the  field through dynamo effects.

A very simple approach to combine magnetic field with the theory of fluid dynamics  is to use the ideal MHD equations. The latter extend the hydrodynamic Euler equations by adding the Lorentz force to the momentum equation and by introducing the induction equation, which describes the temporal evolution of the magnetic field. Because of its simplicity ideal MHD is often used in numerical studies, for example in simulations of large cosmological boxes \citep[e.g.][]{marinacci2015large,marinacci2015effects, dolag2016sz, marinacci2018first}, galaxy clusters \citep[e.g.][]{dolag1999sph, dolag2002evolution}, the magnetic field within individual galaxies \citep[e. e.g.][]{pakmor2014magnetic, pakmor2017magnetic}, the interstellar medium \citep[e.g.][]{kim2018numerical, simpson2023cosmic}, molecular clouds \citep[e.g.][]{grudic2021starforge} and the collapse of protostellar cores \citep[e.g.][]{mellon2008magnetic}.

The ideal MHD equations contain however several assumptions that can be violated, in which case modifications to the underlying equations are required. This is the case, for example, in weakly ionised regions where collisions between neutral and charged particles become rare enough to allow drift velocities between different species. In the low frequency limit one can still derive single fluid equations describing the evolution of the bulk fluid \citep[for a detailed discussion see][]{Pandey2008}. If we neglect the pressure terms in the induction equation, we find three new terms in the induction equation: The resistive Ohmic and ambipolar diffusion and the Hall effect. While the first two terms arise from collisions of charged particles that allow the magnetic field to diffuse, the Hall term becomes important in situations where ions are decoupled from the magnetic field while electrons are still coupled to it. In weakly ionised systems, decoupling can occur because  collisions between neutrals decouple ions from the magnetic field more easily than electrons. Decoupling can also occur in the high frequency limit, when ions cannot follow magnetic fluctuations due to their lower cyclotron frequency compared to electrons. For a full discussion of both regimes see \cite{Pandey2008}. Unlike the other two effects, which only damp MHD waves, the Hall effect introduces two new waves, the whistler (electron-cyclotron) and the ion-cyclotron wave.

A prototypical example of a system in which all three non-ideal MHD effects play a significant role is the collapse of a protostellar core and the subsequent evolution of the weakly ionised protostellar disk. In the ideal MHD approximation, magnetic fields can efficiently transport angular momentum outwards during collapse and prevent disk formation in this so-called magnetic braking catastrophe \citep[see e.g.][]{Allen2003}.  Non-ideal effects can weaken the magnetic field and reduce the efficiency of angular momentum transport, which supports the formation of rotationally supported discs \citep{wurster2016can}. For a recent review of the influence of magnetic fields on the formation of protostellar disks, see \cite{wurster2018role}. 

In the ideal MHD approximation, the newly formed disk can become unstable to the magnetorotational instability \citep[MRI, ][]{velikhov1959stability, chandrasekhar1960stability, fricke1969stability, balbus1991powerful}, which creates turbulence that can act as an effective viscosity, allowing the outward transport of angular momentum and thus accretion onto the central object \citep{shakura1973black,lynden1974evolution}. The MRI can be suppressed by the diffusive non-ideal MHD effects, but the Hall effect can also influence its linear properties, leading to the Hall-shear instability \citep[HSI, ][]{Kunz2008}. Outflows from protostellar disks, which can remove angular momentum, can also be affected by non-ideal MHD effects \citep[e.g.][]{bai2014hall}. Unlike the other two non-ideal effects, the sign of the Hall effect can change, resulting in a bimodal behaviour depending on whether the axes of rotation and the vertical magnetic field are (anti-)aligned. Depending on this condition, magnetic braking can be enhanced or weakened \citep{wurster2021impact}.

Star formation is highly non-linear and arises from the combination of several different physical processes, such as gravity, fluid dynamics or magnetic fields, that all play an important role. Computer simulations are therefore a powerful tool for analysing such systems. Because of their importance, the three non-ideal MHD effects discussed above have been implemented in Eulerian codes such as {\small PLUTO} \citep{lesur2014thanatology}, {\small ATHENA} \citep{bai2014hall}, {\small ATHENA++} \citep{stone2020athena++}, {\small RAMSES} \citep{masson2012incorporating, marchand2018impact}, {\small ZeusTW} \citep{li2011non} and {\small MANCHA3D} \citep{gonzalez2018mhdsts}, as well as Lagrangian particle methods such as the SPH code of \cite{tsukamoto2017impact}, {\small PHANTOM} \citep{price2018phantom} and {\small GIZMO} \citep{hopkins2017anisotropic}. Eulerian codes typically suffer from larger advection in regions with high bulk velocities whereas the popular Lagrangian method called smoothed particle hydrodynamics (SPH) produces noisier results with lower accuracy but offers a density-adaptive resolution.

A relatively new approach that attempts to combine the advantages of a Galilei-invariant Lagrangian method with the high accuracy of the finite volume methods typically used in Eulerian codes is the moving mesh method as implemented in the {\small AREPO} code \citep{springel2010pur,weinberger2020arepo}. It solves the fluid equations on an unstructured, evolving grid, which makes it particularly interesting for the analysis of systems with large density gradients, due to its adaptive cell sizes, and systems with large bulk velocities, such as rotationally supported disks. As we have shown in \cite{ZierMRI} and \cite{ZierGravitoturbulence2023} the moving-mesh approach is indeed able to resolve the MRI as well as the gravitational instability (GI) with high accuracy in the shearing box approximation \citep{hill1878collected,goldreich1965ii, zier2022simulating}.

In \cite{zier2023} (Paper I from now on) we extended the {\small AREPO} code to include resistive Ohmic and ambipolar diffusion. The module is based on a least-squares fit to obtain accurate estimates of the magnetic field gradients and shows comparable accuracy to the state-of-the-art code {\small ATHENA++}, which uses a Eulerian method.
The aim of this paper is to add the Hall effect to our implementation, which is numerically quite challenging as it is known to easily destabilise numerical algorithms that use second-order accurate time integration as in {\small AREPO}. However, we show that the algorithm can be stabilised by locally increasing the magnetic diffusion in a judicious fashion.

The paper is structured as follows. In Section~\ref{sec:theory} we introduce the non-ideal MHD equations and derive the linear dispersion relation of the new MHD waves introduced by the Hall effect. In Section~\ref{sec:numericalMethods} we introduce the moving mesh method and discuss the implementation of the term describing the Hall effect. For discretization we use the same method that we have already used for Ohmic and ambipolar diffusion. In Section~\ref{sec:stabilityHall} we show by simulating a magnetic C-shock that our scheme becomes unstable when the Hall effect is strongly dominant over the magnetic diffusion originating from explicit diffusion or from  numerical resistivity. We also show that by locally increasing the Ohmic diffusion we can stabilize our scheme and find convergence to the semi-analytic solution in this case. In Section~\ref{subsec:PropagationWhistler}, we demonstrate that our new scheme with the additional Ohmic diffusion is able to reproduce the analytical dispersion relation for the whistler and ion-cylcotron waves. We also compare the diffusivity of our scheme with an implementation using the more diffusive HLL Riemann solver instead of an Ohmic diffusion. As a first application, in Section~\ref{subsec:Collapse} we present simulations of the collapse of a magnetised cloud with constant non-ideal MHD parameters. We show that, depending on the sign of the Hall effect, the magnetic braking can be enhanced or weakened. Finally, we summarise our results in Section~\ref{sec:discussionSummary}.

\section{Theory}
\label{sec:theory}
\subsection{The non-ideal MHD equations}
The non-ideal MHD equations can be written as a conservation law:
\begin{equation}
    \frac{\partial \bm U }{\partial t}+ \nabla \cdot \bm F_{\rm ideal}(\bm U) +  \nabla \cdot \bm F_{\rm res}(\bm U) +  \nabla \cdot \bm F_{\rm H}(\bm U) =0.
    \label{eq:nonidealMHDEquations}
\end{equation}
Here, we introduced the state vector $\bm U$, the ideal MHD flux function $\bm F_{\rm ideal}$, the flux function $\bm F_{\rm res}$ for the non-ideal MHD effects of Ohmic and ambipolar resistivity, and the flux function $\bm F_{\rm H}$ describing the Hall effect. They are given in Heaviside-Lorentz units by:
\begin{align}
\bm U  = \begin{pmatrix}
   \rho \\
   \rho \bm v  \\
   \rho e  \\
   \bm B\\
   \end{pmatrix},  \;\;\;\;\;\; 
   \bm F_{\rm ideal}(\bm U) =  \begin{pmatrix}
   \rho \bm v\\
   \rho \bm v \bm v^T + P -\bm B \bm B^T\\
   \rho e  \bm v + P \bm v -\bm B(\bm v \cdot \bm B)\\
   \bm B \bm v^T -\bm v \bm B^T
   \end{pmatrix}, \\
\bm F_{\rm res}  = \begin{pmatrix}
   0 \\
   0 \\
   \eta_{\rm OR} \nabla \bm B  + \eta_{\rm AD}\left[ \left(\bm J\times \bm b\right) \bm b - \bm b \left(\bm J\times \bm b\right)\right] \\
   \eta_{\rm OR} \bm J \times \bm B +\eta_{\rm AD} \left\{\left[\left(\bm J \times  \bm B \right) \times \bm b \right] \times \bm b \right\}\\
   \end{pmatrix}, \\
   \bm F_{\rm H}  = \begin{pmatrix}
   0 \\
   0 \\
   \eta_{\rm H} \left(\bm J\bm b - \bm b \bm J\right) \\
  0\\
   \end{pmatrix}, 
   \label{eq:nonidelMHDSourceTerms}
\end{align}
where $\rho$, $\bm v$, $e$, $\bm B$, $\bm b = \bm B / \left|\bm B\right|$, $P$, and $\bm J = \nabla \times \bm B$ are the density, velocity, total energy per unit mass, magnetic field strength, direction of the magnetic field, pressure, and electric current, respectively. The specific energy $e = u + \frac{1}{2} \bm v^2 + \frac{1}{2 \rho} \bm B^2$ consists of the thermal energy  per mass $u$, the kinetic energy density $\frac{1}{2} \bm v^2$, and the magnetic field energy density $\frac{1}{2 \rho} \bm B^2$. The pressure $P=p_{\rm gas} + \frac{1}{2}\bm B^2$ includes a thermal and a magnetic component. The system of equations is closed by the equation of state (EOS), which expresses $p_{\rm gas}$ as a function of the other thermodynamical quantities. In this paper, we mostly use an isothermal EOS,
\begin{equation}
     p_{\rm gas} = \rho c_s^2,
\end{equation}
with constant isothermal sound speed $c_s$.
In some cases we also use an adiabatic equation of state
\begin{equation}
     p_{\rm gas} = \left(\gamma -1 \right) \rho u,
\end{equation}
where we introduced the adiabatic coefficient $\gamma = 5/3$.
The diffusivities $\eta_{\rm OR}$, $\eta_{\rm AD}$ and $\eta_{\rm H}$ describe the strength of the Ohmic and ambipolar diffusion and the Hall effect, respectively, which are in general a function of the magnetic field strength and the local chemical composition. In \cite{zier2023} (Paper I) we discussed the resistive term $\bm F_{\rm res}$ and in this paper we will concentrate on the Hall term $\bm F_{\rm H}$.

\subsection{Linear MHD waves in Hall MHD}
\label{subsec:linearWaveMHDHall}

As one can see in equation (\ref{eq:nonidelMHDSourceTerms}), the Hall effect does not affect the energy equation but only leads to a change in the topology of the magnetic field. In contrast to the other two non-ideal MHD effects it is not diffusive but dispersive, which has not only an impact on its physical nature but also on the stability of numerical schemes. This can be seen by assuming a system with a purely vertical magnetic field $\bm B = B \hat{e}_z$, and by adding perturbations $\delta v_x$, $\delta v_y$, $\delta b_x$, and $\delta b_y$ to the velocity and magnetic field, which are travelling in the $z$-direction, i.e.~they are proportional to $e^{i( \omega t -  k z)}$. If one only takes into account the induction equation with the Hall effect, and the momentum equation with the Lorentz force, one ends up with the linearized equations \citep{marchand2018impact}:
\begin{gather}
    \rho i \omega \delta v_x = - i k B \delta b_x,\\
    \rho i \omega \delta v_y = - i k B \delta b_y,\\
    i \omega \delta b_x = i k B \delta v_x + \eta_{\rm H} k^2 \delta b_y,\\
    i \omega \delta b_y = i k B \delta v_y - \eta_{\rm H} k^2 \delta b_x.
\end{gather}
This system only has a non-trivial solution if the dispersion relation
\begin{equation}
    \omega^4 - \omega^2 \left(2k^2 v_A^2 + \eta_{\rm H}^2 k^4\right) + k^4 v_A^4 = 0,
\end{equation}
with the Alfvén velocity $v_A = B/\sqrt{\rho}$ is fulfilled. It has two solutions for a positive frequency $\omega$ \citep{Balbus2001}:
\begin{equation}
    \omega_{1,2}= \pm \frac{\eta_H k^2}{2} + \sqrt{\left(\frac{\eta_{\rm H} k^2}{2}\right)^2 + k^2 v_A^2},
    \label{eq:dispersionRelationWhistler}
\end{equation}
which simplify to the standard Alfvén waves for $\eta_{\rm H} = 0$. These new waves are the whistler wave ($+\,{\rm sign}$) and the ion cyclotron wave ($-\,{\rm sign}$), and their velocities are given by:
\begin{equation}
    c_{1,2} = \frac{\omega}{k} = \pm \frac{\eta_{\rm H} k}{2} + \sqrt{\left(\frac{\eta_{\rm H} k}{2}\right)^2 + v_A^2},
\end{equation}
which shows that the Hall effect is dispersive. 

One of the waves always travels faster than the Alfvén speed, and for small perturbations (large $k$) its velocity is even unbound. This unphysical behaviour is a direct consequence of neglecting the electron inertia and assuming effectively an infinite speed of light (charge neutrality). By using a two-fluid model one can show that the frequency of whistler waves is actually bounded by the electron cyclotron frequency \citep{Srinivasan2011}. From a numerical point of view the unbounded velocity is also problematic since this means that perturbations on the grid scale (including unphysical noise) travel the fastest, which can make numerical schemes unstable. In pseudo-spectral codes that solve the system in Fourier space short-range wavelengths can be easily removed, and they therefore show better stability behaviour. In fact, \cite{Kunz2013} demonstrate that their pseudo-spectral algorithm is stable if one uses a third-order time integration scheme, while \cite{falle2003} showed that first and second order explicit schemes are unconditionally unstable. Methods that integrate in real space and only use a second-order accurate time integration scheme therefore require some kind of dissipation which  efficiently damps small scale whistler waves. This can be a physical (Ohmic or ambipolar diffusion), an artificial but explicit (e.g.~artificial resistivity in SPH) or an implicit numerical diffusion (e.g.~through the Riemann solver).

\section{Implementation of the Hall effect on a moving mesh}
\label{sec:numericalMethods}

\subsection{Ideal MHD in AREPO}

In this paper, we use the moving mesh code {\small ARPEO} \citep{springel2010pur, weinberger2020arepo} to solve equation (\ref{eq:nonidealMHDEquations}) on a moving, unstructured Voronoi mesh with the finite volume method using the Riemann HLLD solver \citep{miyoshi2005multi}. \cite{springel2010pur} introduced the code and described the general algorithms for mesh construction and hydrodynamic simulations. \cite{pakmor2011magnetohydrodynamics} and \cite{pakmor2013simulations} extended it to ideal MHD, and \cite{pakmor2016improving} introduced a new gradient estimate and time integration to achieve in most test problems second-order convergence in time and space. \cite{zier2022simulating} further improved the method by introducing a higher-order flux integration that reduces the production of noise on the grid scale, which is especially useful in shear flows.

The continuum equations (\ref{eq:nonidealMHDEquations}) conserve the condition $\nabla \cdot \bm B = 0$, but this property can get lost once the equations are discretized.  This can in turn lead to purely numerical instabilities and therefore to large deviations from the $\nabla \cdot \bm B = 0$ condition. \cite{pakmor2013simulations} introduced the Powell scheme \citep{powell1999solution} in {\small ARPEO} for convergence control that diffuses the magnetic monopoles whereas \cite{pakmor2011magnetohydrodynamics} implemented the Dedner scheme \citep{dedner2002hyperbolic} that additionally damps it. In this paper, we will use the former approach in all simulations.

We note that static grid codes often use the constrained transport method that is able to conserve the condition $\nabla \cdot \bm B = 0$ up to machine precision even for the discretized equations \citep{evans1988simulation}. Implementing this method for dynamic meshes becomes very complicated, however, and also tends to be more diffusive \citep{mocz2014constrained}.

\subsection{Magnetic diffusion in {\small ARPEO}}
In Paper~I we introduced a new diffusion solver to integrate the resistive flux function $F_{\rm res}$ in equation (\ref{eq:nonidealMHDEquations}) using operator splitting and the finite volume method.  We use the same basic approach also for the Hall effect. By integrating equation (\ref{eq:nonidealMHDEquations}) over the volume $V_i$ of a Voronoi cell $i$ we find for the non-ideal part:
    \begin{equation}
\frac{\partial \bm Q_i}{\partial t} = - \int_{V_i} \nabla \cdot \left(\bm F_{\rm res} + \bm F_{\rm H}\right)= - \int_{\partial V_i} \hat{n} \cdot \left(\bm F_{\rm res} + \bm F_{\rm H}\right).
\end{equation}
Here we introduced the vector of conserved quantities $\bm Q_i = \int_{V_i} \bm U$ and applied Gauss' theorem. For the time integral we use the standard second-order accurate Runge-Kutta scheme while the surface integral can be rewritten as a sum over all interfaces of one cell. The integral over one interface can be approximated by evaluating the Flux function at the center of the interface and by multiplying it with its area. We note that in \cite{zier2022simulating} we showed that this approximation can become inaccurate for the ideal MHD flux in shear flows, where a higher order approximation is required. For the non-ideal MHD terms we did not find a significant improvement by using those more expensive integration rules and therefore we will not use them here. For the integration of the non-ideal terms we use again the Strang-splitting scheme.

To apply the Hall term we are left with evaluating $\bm F_{\rm H}$ at the interface, which requires according to equation (\ref{eq:nonidelMHDSourceTerms}) accurate estimates for the magnetic field and its gradients. In Paper~I we introduced a method based on performing a least square fit at the centre of the interface, taking into account the values of the magnetic field at all neighbouring cells. We will use the same method in this paper and therefore refer to Paper~I for full details.

\subsection{Time step constraints}

The Hall effect leads to an additional time step constraint which can be written as \citep{bai2014hall}:
\begin{equation}
\Delta t_{\rm H} = C_{\rm H}\frac{\Delta x_i^2}{2 d\,\eta_{\rm H}}, \label{eq:timeStepConstrainedHall}
\end{equation}
where $\Delta x_i = \left[ 3V_i / \left(4 \pi\right) \right]^{1/3}$ is the effective radius  of the Voronoi cell $i$ with volume $V_i$ in three dimensions ($\sqrt{V_i /\pi}$ in two dimensions) and $d$ is the number of dimensions. $C_{\rm H}$ is the Courant number for the Hall effect which we typically set to $0.5$. Like the other non-ideal MHD terms, the Hall effect leads for high resolutions to a very restrictive time step constraint. In contrast to the other two effects the Hall effect cannot be accelerated by using super-time stepping (see \cite{berlok2020braginskii} for an implementation in {\small AREPO} for Braginskii viscosity), so the only option would be subcycling. In this paper we will not use such methods but rather observe the time step constraint, which is not yet a problem at the only moderately high resolution we employ.

\subsection{The strength of the Hall effect}

The strength of the Hall effect can be parameterised by the diffusion coefficient $\eta_{\rm H}$. It strongly depends on the local chemical composition and magnetic field strength. As a consequence factors such as the cosmic ray ionization rate and the dust size distribution can have a large impact on $\eta_{\rm H}$. In contrast to the coefficients for Ohmic and ambipolar diffusion, the Hall coefficient can be both positive and negative, and will generally have different signs throughout the simulation domain. Although there exist library solutions \citep[e.g.][]{wurster2016nicil} that allow the calculation of the non-ideal MHD diffusion coefficients in equilibrium the values of $\eta_{\rm H}$ can vary over several orders of magnitudes. It is therefore difficult to anticipate when and where the Hall effect is dominating \citep{wurster2018role}.

The goal of this paper is to demonstrate the accuracy of our non-ideal MHD solver, and therefore we will use only constant or parameterized coefficients:
\begin{equation}
    \eta_{\rm H} = H \left|B\right|,
\end{equation}
where $H$ is a constant, for simplicity and definiteness. The Hall effect dominates over the ideal MHD induction term if ${\eta_{\rm H} k}/{2} \gg v_A$ holds, which can be translated to a critical length scale $l_{\rm H} = {\eta_{\rm H}}/{v_A}$. This length scale allows us to define the Lundquist number
\begin{equation}
    L_{\rm H}^* = \frac{1}{l_{\rm H} k} = \frac{v_A}{\eta_{\rm H} k}.
    \label{eq:lundquistNumber}
\end{equation}
 for perturbations with wave vector $k$.

\section{Stabilizing the Hall term in the Hall-dominated regime}
\label{sec:stabilityHall}

As we have discussed in Section~\ref{subsec:linearWaveMHDHall}, the Hall-MHD equations without resistivity allow for whistler waves whose velocity becomes arbitrarily large for small wavelengths. This can destabilize our scheme since numerical noise that is inherent in simulations on a moving mesh \citep[see e.g.~][]{zier2022simulating} travels the fastest.  In fact, we found that our scheme becomes unstable in simulations of a gravitational collapse of a protostellar core if we only take into account the Hall effect and no diffusion.  The instability can be observed first in the densest region, which is equivalent to the region with the highest spatial resolution. This behaviour can be understood because the numerical diffusion, which always exists in our scheme, decreases with better spatial resolution, which in turn means that  its stabilizing influence on  the Hall effect diminishes in the core. We thus expect that it first falls there below the stability threshold for the Hall effect.

\cite{Kunz2013} show that for a third-order time integration scheme no additional diffusion would be required, but this is complicated to implement in a multi-physics code such as {\small AREPO}. In the literature different methods were therefore introduced to stabilize the Hall effect if there is not enough physical resistivity but a second-order accurate time integration scheme is still kept. \cite{osullivan2006halldiffusion} proposed a semi-implicit integration method, which is marginally stable. In this scheme, one Cartesian component of the magnetic field is updated with a fully explicit timestep. The second component uses this updated field for a semi-implicit step. Finally, the last step for the third component is fully implicit, using the two already updated components. The generalization to 3D was performed in  \cite{osullivan2007halldiffusion3D} and  implemented into {\small ATHENA} by \cite{bai2014hall}. 

\cite{Bai2017} found that in spherical coordinates this scheme becomes unstable and therefore used instead a modified HLL Riemann solver that takes into account the whistler waves. The HLL Riemann solver is significantly more diffusive but its increased numerical resistivity helps to stabilize the Hall effect. \cite{marchand2019} showed that in simulations of the collapse of protostellar cores, the modified HLL solver can lead to the artificial creation of angular momentum if the whistler wave is accounted for in all variables. We will show in Section~\ref{subsec:waveDiffusivity} and Appendix~\ref{app:RiemannColappse} that the HLL Riemann solver is indeed more diffusive, and that this can reduce the numerical accuracy even in regions where the Hall effect is subdominant.

Another option to stabilize the Hall effect is to use an artificial resistivity, which is anyway required in SPH simulations. One can also introduce a hyper-resistivity, which is very efficient in damping noise on small scales but does not significantly affect larger scales \citep{Srinivasan2011}. This would require the calculation of higher derivatives of the magnetic field, which is however very expensive and not particularly accurate on an unstructured mesh. As we will discuss in the following we employ instead a standard artificial resistivity.

 \begin{figure*}
    \centering
    \includegraphics[width=1\linewidth]{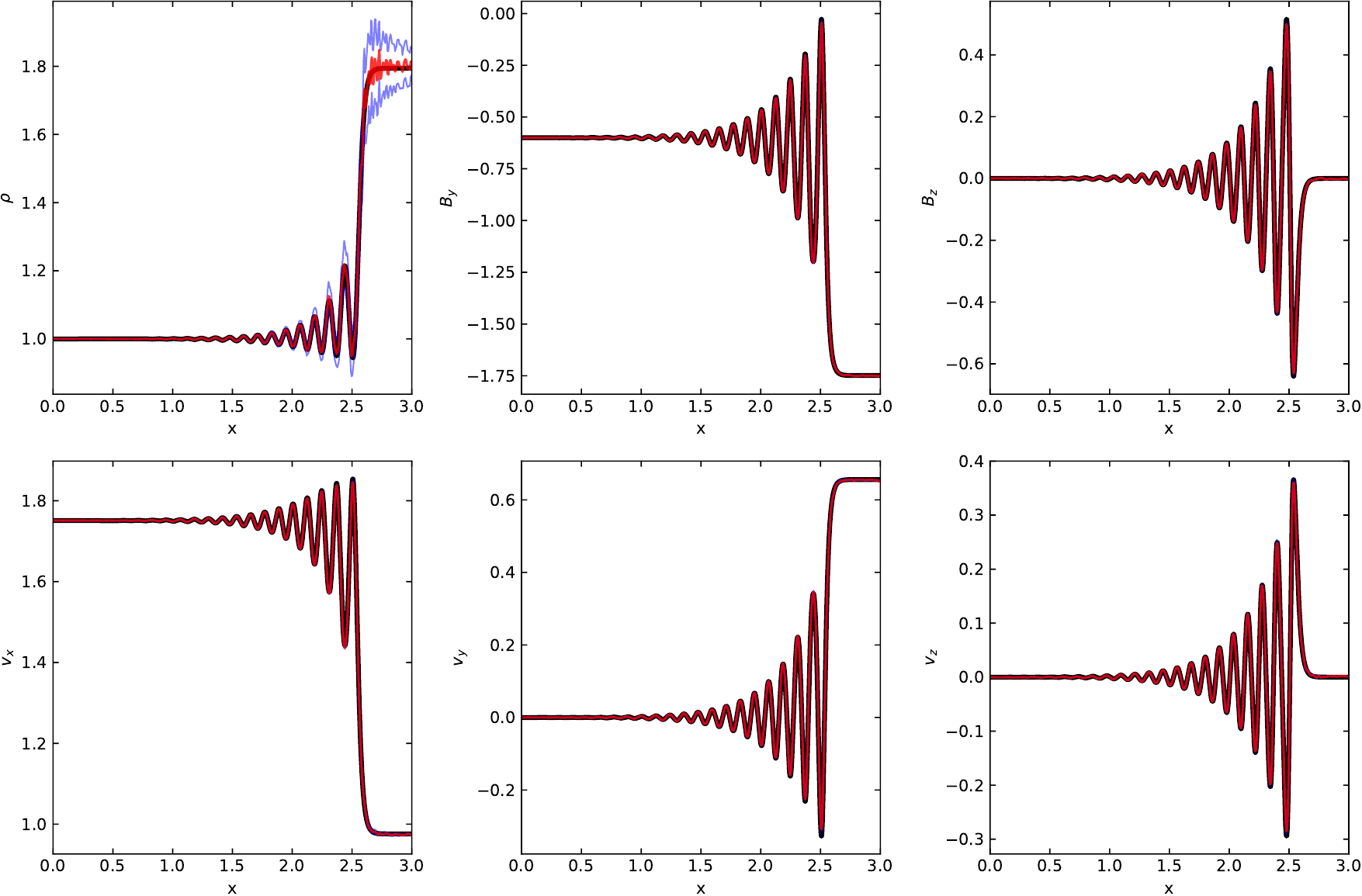}
    \caption{The structure of the C-shock at $t=5$ for a simulation with physical resistivity $\eta_{\rm OR} = 0.001$ and $N=160$.
    The black line corresponds to the best fit semi-analytically expected structure of the shock with $\eta_{\rm OR} = 0.0012$, while the red line shows the average values in a simulation with {\small AREPO} at time $t=5$. The blue lines show the  $\pm 1$ standard deviation in each bin.}
    \label{fig:CShockHallWithDiff20Res160Continous}
\end{figure*}

\subsection{The stability of a C-shock}
\begin{table}
    \centering
    \begin{tabular}{c|c}
    \hline
        $\eta_{\rm OR}$ & $L_x$\\
        \hline
       0.005 & 1\\
       0.002 &2\\
        0.001   &3\\
         0.0005  & 5\\
        0.0002  &12\\
        \hline
    \end{tabular}
    \caption{The box size for simulations of a C-shock with explicit diffusion $\eta_{\rm OR}$.}
    \label{tab:boxSizeExplicitDiffusion}
\end{table}

\begin{table}
    \centering
    \begin{tabular}{c|c|c|c|c|c}
    \hline
        Variable & $\rho$ & $v_x$ & $v_y$ & $B_x$ & $B_y$ \\
        \hline
        Pre-shock values & 1 & 1.751& 0& -1 & -0.6\\
        Post-shock values & 1.7942 &	0.9759 & 0.6561 & -1 & -1.74885\\
        \hline
    \end{tabular}
    \caption{Initial conditions for the C-shock tests with sound speed $c_s = 0.1$.}
    \label{tab:CShockInitialConditions}
\end{table}

In ideal MHD, shocks contain discontinuous jumps in the parallel component of the magnetic field and in the hydrodynamic quantities. The non-ideal MHD terms smooth these discontinuities, making them continuous \citep[``C-type'' shock,][]{Draine1980}. As we have shown in Paper~I, Ohmic and ambipolar diffusion lead to a pure smoothing of the jump, and all quantities stay monotonous functions in space in the direction normal to the shock front.  This changes for the Hall effect, which introduces oscillations close to the shock front. These oscillations have to be damped by Ohmic or ambipolar diffusion, otherwise they would continue in the steady state throughout the whole system. This  can be seen by solving for the static profile of a C-shock, which we further discuss in Appendix~\ref{app:CShock}. The smaller the explicit diffusion is compared to the Hall effect, the more oscillations can be observed. In simulations lacking any explicit diffusion the oscillations will nevertheless be damped by numerical diffusion $\eta_{\rm N}$, making the C-shock an ideal system to measure $\eta_{\rm N}$ but also for investigating the numerical stability of the Hall term.

We use a similar setup as in Paper~I, which means we will simulate a box of size $L_x \times L_y \times L_z$ with $L_y = L_z$. The shock is travelling in the $x$-direction and we use periodic boundary conditions in the  $y$- and $z$-directions. In the $x$-direction we extend the box by $0.5$ on both sides and set for $-0.5 < x < 0$ all primitive variables to the pre-shock values, and for $L_x < x < L_x +0.5$ to the post-shock values. We use for the extended box periodic boundary conditions, which ensures that the total number of cells stays constant. With a fixed box size it becomes expensive to perform resolution studies, since to double the spatial resolution in the $x$-direction we would have to use 8 times the amount of cells. To circumvent this problem we use a box size $L_y = 10 / N$, where $N^3$ is the desired number of cells within a unit box. For strong Ohmic diffusion we can use $L_x = 1$ but for simulations with weaker diffusion the shock becomes more extended and we have to increase $L_x$. We present the box sizes for simulations with explicit diffusion in \cref{tab:boxSizeExplicitDiffusion}. In \cref{tab:CShockInitialConditions} we give the pre- and post-shock values for our simulations. We use $\eta_{\rm H} = 0.02 \left| B\right| = H \left| B\right|$, and as we show in Appendix~\ref{app:CShock} the structure is fully determined by the ratio $\eta_{\rm OR} / H$. The setup is inspired by \cite{marchand2018impact}, and we use an isothermal equation of state with sound speed $c_s = 0.1$.

We run all simulations until $t=5$ and afterwards bin the Voronoi cells in the $x$-direction. We also calculate for each bin the standard deviation as a measure of the noise in our simulations, which is only significant in the density profile. We use $2 N$ bins per unit length which means that on average 50 cells can be found in each bin. Afterwards, we calculate profiles for different $\eta_{\rm OR}$ and try to fit the profile that minimizes the 
\begin{equation}
   L_1 = \frac{1}{N_{\rm bin}} \sum_{i =1}^{N_{\rm bin}}\left| \bar{B}_{z,i,s} - B_{z,i,t}\right|,
   \label{eq:definitionL1}
\end{equation} error for $B_z$. We introduced here the average magnetic field $\bar{B}_{z,i,s}$ of our simulations in the $z$-direction in each bin, and the value at the center of each bin for the semi-analytic profile $B_{z,i,t}$.

For the initial conditions we use two different setups: In the first one we set the primitive variables to the pre-shock value for $x < x_0$ and to the post-shock value for $x_0 <x$. In the second one we smooth the initial conditions by using the equation
\begin{equation}
    Y\left(x\right) = \frac{Y_l +Y_r}{2} + \frac{Y_r -Y_l}{2} \tanh \left[5 \left(x-x_0\right)\right]
\end{equation}
for the primitive variable $Y$, where the subscript $l$ refers to the pre-shock value and $r$ the post-shock value. We typically use $x_0 = L_x - 0.5$ as the initial position of the shock. As in Paper~I we construct initially an unstructed mesh by replicating a box with $10^3$ cells.

We perform simulations with explicit diffusion for which we expect convergence to the semi-analytical profile and simulations without explicit diffusion, which allows us to measure $\eta_N$ and the maximum ratio $\eta_{\rm H} /\eta_{\rm OR}$ for which our scheme is stable.

\subsubsection{With explicit diffusion}
\label{subsubsec:CshockExplicit}

As one can see in  \cref{fig:CShockHallWithDiff20Res160Continous} our method is able to accurately resolve the structure of a C-shock even in the strongly Hall-dominated regime. The largest deviations are found in the density profile, which is also the only quantity showing significant noise. As we have already discussed in Paper~I this noise can also be observed without non-ideal MHD and therefore is not caused by inaccuracies in our non-ideal MHD solver. 

\begin{figure}
    \centering
    \includegraphics[width=1\linewidth]{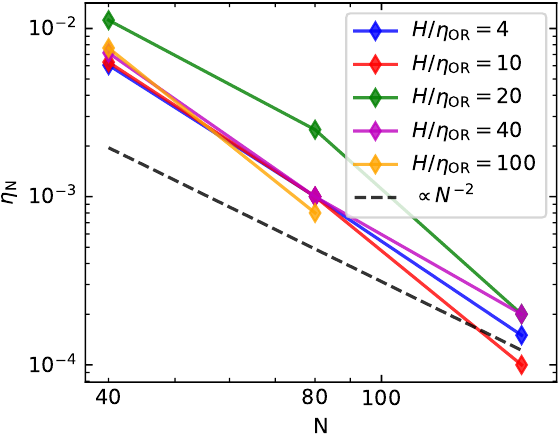}
    \caption{The numerical resistivity $\eta_{\rm N}$ as a function of resolution for different simulations of the C-shock with physical resistivity $\eta_{\rm OR}$. We show the averaged values for continuous and discontinuous initial conditions.}
    \label{fig:numerical_res_with_diff}
\end{figure}

We performed a parameter study in which we varied $\eta_{\rm OR}$ and the resolution, for which we show the most important parameters in  \cref{tab:OverViewCShockExplicit}. The numerical resistivity can be defined as the difference between the measured one and the explicit one we set at the beginning of the simulation. As expected it strongly decreases with increasing resolution, and we find convergence to the physical resistivity. This can also be seen in \cref{fig:numerical_res_with_diff}, where we find a similar scaling of $\eta_N$ with the resolution $N$. But we also observe that the trend cannot be described well by a $N^{-2}$ power law. For $\eta_{\rm OR}/ H \geq 0.1$, all simulations stay stable whereas for smaller explicit diffusivity some simulations with high resolution become unstable. We observe that simulations with discontinuous initial conditions become immediately unstable at the beginning of the calculation while simulations with continuous initial conditions first form a C-shock profile before they become unstable.

\begin{table}
    \centering
    \begin{tabular}{c|c|c|c|c|c}
    \hline
        Res. N & $\eta_{\rm OR}$ & $\eta_{\rm tot, D}$ &  $\eta_{\rm tot, C}$ & $L_{\rm 1,D}$ & $L_{\rm 1,C}$\\
        \hline
        40 & 0.005 & 0.0106 &0.0101 & 0.0060 & 0.0058\\
        80 & 0.005 &0.0059 & 0.0061& 0.0030 &0.0029\\
        160 & 0.005 & 0.0052 &0.0051&0.0009 &0.0010\\
   
        40 & 0.002 &0.0085 & 0.0085 & 0.0046 & 0.0046\\
        80 & 0.002 &0.0030 & 0.0030 & 0.0027 & 0.0027\\
        160 & 0.002&0.0021 & 0.0021& 0.0018 & 0.0018 \\
  320& 0.002 & 0.0020 & 0.0020 & 0.0004 & 0.0003\\
                40 & 0.001 & 0.0123 &0.0121  &0.0033 &0.0033\\
        80 & 0.001 & 0.0035 &0.0035 &0.0042 & 0.0040\\
        160 & 0.001 & 0.0012 &0.0012 &0.0022 & 0.0020\\
        320& 0.001 & unstable & unstable&\\
                40 & 0.0005 &0.0075 & 0.0078 & 0.0023 & 0.0022\\
        80 & 0.0005 &0.0015 &0.0015 &0.0039 & 0.0040\\
        160 & 0.0005 &0.0007 & 0.0007 & 0.0122 & 0.0105\\
        320& 0.0005 & unstable & unstable\\
        40& 0.0002 &0.0078 & 0.0079& 0.0014&0.0014\\
        80& 0.0002 & 0.0010 &0.0010 & 0.0032& 0.0028\\
        120& 0.0002 & unstable &unstable &\\
        160& 0.0002 & unstable & unstable &\\
        \hline
    \end{tabular}
    \caption{Overview of simulations of a C-shock with explicit Ohmic resistivity $\eta_{\rm OR}$. We use on average $N$ cells per unit length in the $x$-direction and measure the total resistivity $\eta_{\rm tot}$ by fitting the semi-analytic profile to the simulation data.  We performed simulations with discontinuous (D) and continuous (C) initial conditions, and also give the final $L_1$ error for $B_z$, which we try to minimize.}
    \label{tab:OverViewCShockExplicit}
\end{table}

\subsubsection{Without explicit diffusion}

\begin{figure*}
    \centering
    \includegraphics[width=1\linewidth]{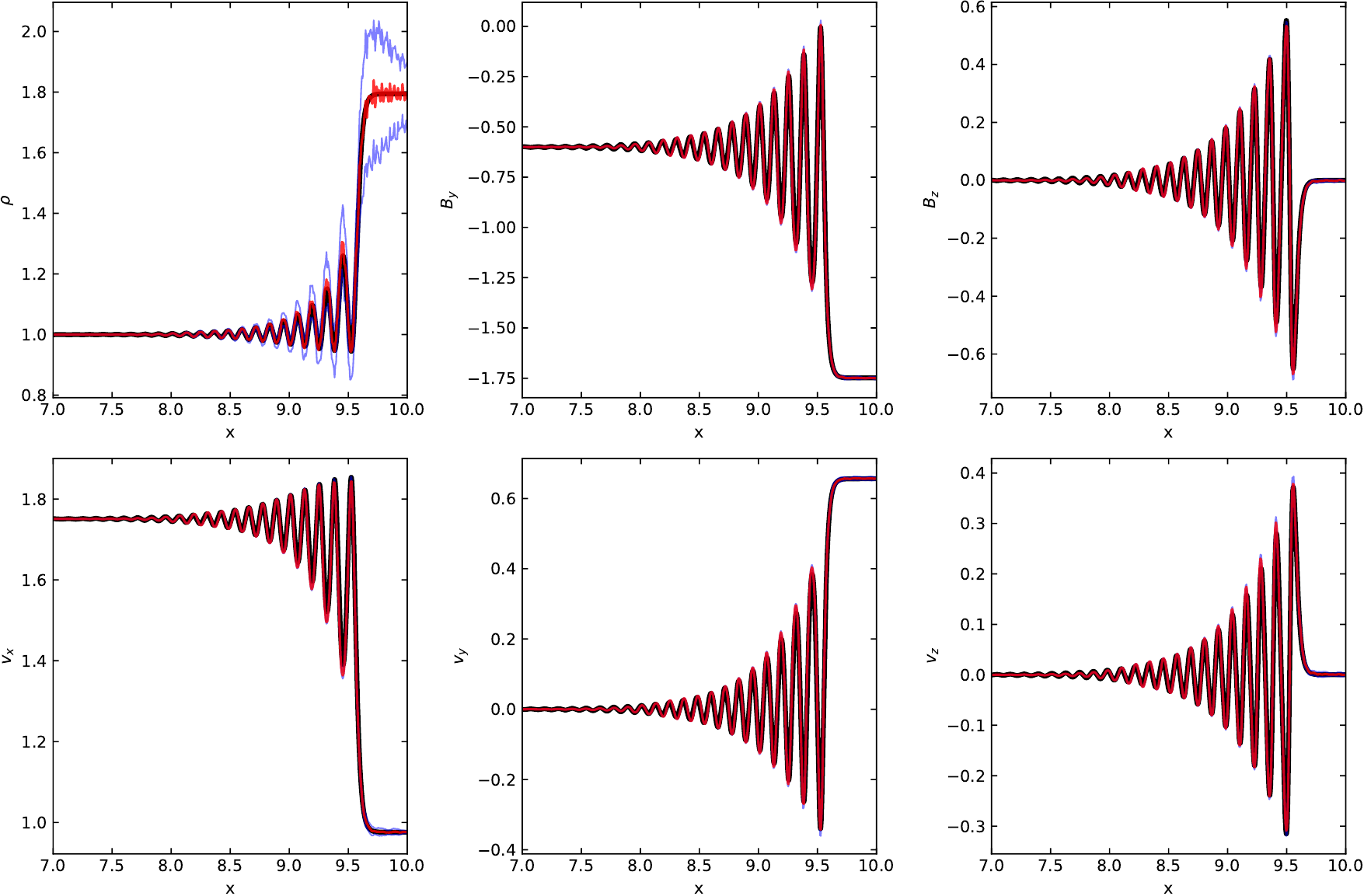}
    \caption{The structure of the C-shock at $t=5$ for a simulation without physical resistivity, $N=80$ and continuous initial conditions.
    The black line corresponds to the best fit semi-analytically expected structure of the shock with $\eta_{\rm OR} = 0.0009$, while the red line shows the average values in a simulation with {\small AREPO} at time $t=5$. The blue lines show the  $\pm 1$ standard deviation in each bin}
    \label{fig:HallNoDiff80Disc}
\end{figure*}

\begin{figure}
    \centering
    \includegraphics[width=1\linewidth]{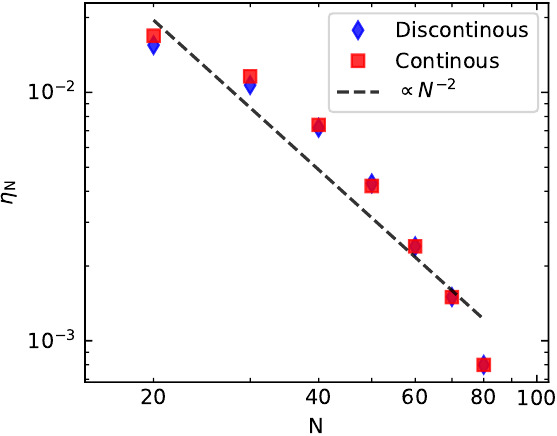}
    \caption{The numerical resistivity $\eta_N$ as a function of resolution for simulations of the C-shock without physical resistivity. 
    We show the results for continuous and discointous inital conditions.}
    \label{fig:numerical_res_no_diff}
\end{figure}

To ensure that the full C-shock can be resolved in simulations without explicit diffusion, we use in this section a fixed box size $L_x = 10$. As one can see in \cref{fig:HallNoDiff80Disc}, the numerical resistivity acts indeed similarly to a physical one which allows us to measure $\eta_N$ by close inspection of the shock profile. In \cref{tab:OverViewCShockWithout} we give an overview of all simulations we performed without physical resistivity. $\eta_N$ is more or less independent of the initial conditions and decreases with increasing resolution.  For $N=90$ the simulations are still stable but we can observe larger noise especially in the density, which is also reflected in the strongly increased $L_1$ error compared to $N=80$. For even higher resolution the simulations become unstable, so we conclude that $N=90$ is close to the instability. In \cref{fig:numerical_res_no_diff} we show the dependence of $\eta_N$ on the resolution. We find that $\eta_N$ does not follow a simple power law, which prevents us from defining a simple lower boundary for the numerical resistivity.

\begin{table}
    \centering
    \begin{tabular}{c|c|c|c|c|c}
    \hline
        Res N & $\eta_{\rm N, D}$& $\eta_{\rm N, C}$  & $L_{\rm 1,D}$ & $L_{\rm 1,C}$\\
        \hline
        20 &0.0155 & 0.0169 &0.0009 & 0.0008\\
        30 &0.0107& 0.0116 &0.0010& 0.0010\\
        40 &0.0072  &0.0074 & 0.0012 & 0.0011\\
        50 & 0.0043 &0.0042 & 0.0013 & 0.0013\\
        60 &0.0024 & 0.0024 & 0.0018 &0.0017\\
        70 & 0.0015& 0.0015 & 0.0025 & 0.0024\\
        80 & 0.0008 & 0.0009 & 0.0027 & 0.0028\\
        90 & 0.0008& 0.0009 & 0.0070&0.0091\\
        100 &unstable &unstable  &- &-\\
        120 & unstable & unstable &- &-\\
     \hline

    \end{tabular}
    \caption{Overview of simulations of a C-shock without explicit Ohmic resistivity. We use on average $N$ cells per unit length in the $x$-direction and measure the numerical resistivity $\eta_{\rm N}$ by fitting the semi-analytic profile to the simulation data. The simulation can be stable or unstable for discontinous (D) or continous (C) initial conditions. We use a box size $L_x = 10$ to make sure that the boundary does not affect the results and give the average $L_1$ error for $B_z$ at $t=5$.}
    \label{tab:OverViewCShockWithout}
\end{table}

\subsubsection{Convergence}
\label{subsec:convergenceCShock}

\begin{figure}
    \centering
    \includegraphics[width=1\linewidth]{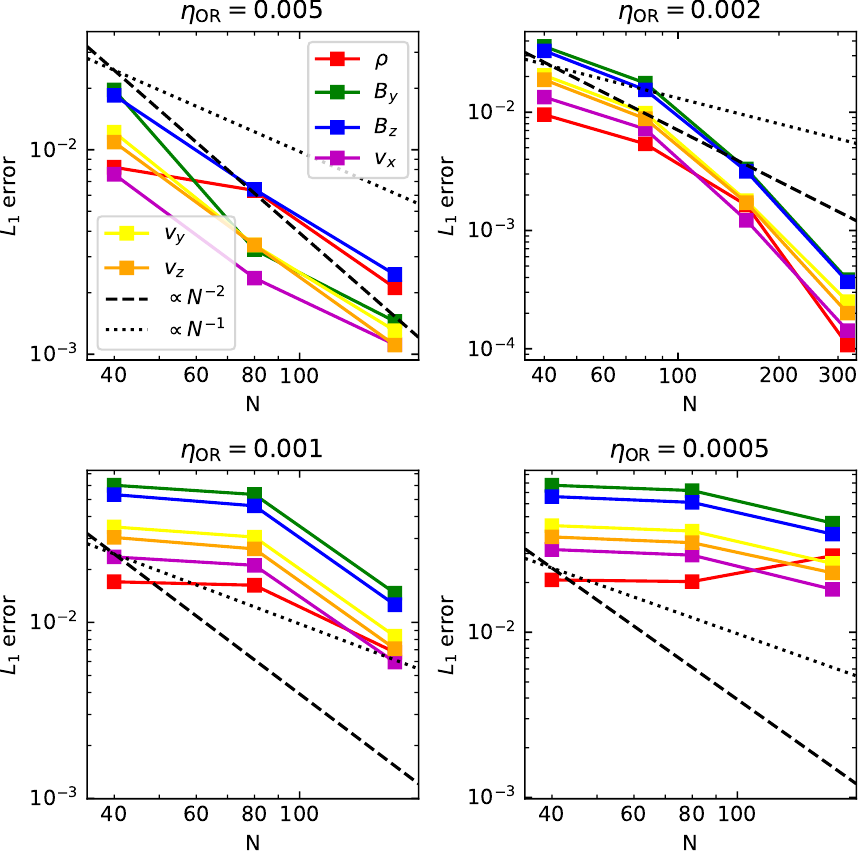}
    \caption{The $L_1$ error as a function of resolution for the C-shock test with different physical resistivity $\eta_{\rm OR}$. More details can be found in Section~\ref{subsec:convergenceCShock}.}
    \label{fig:L1_error_CShock_With_Diffusion}
\end{figure}

To verify the convergence of our method we define the $L_1$ error norm for each quantity at time $t=5$  following equation~(\ref{eq:definitionL1}), and we compare our profiles directly with the semi-analytic result for the same physical resistivity. As one can see in \cref{fig:L1_error_CShock_With_Diffusion} we find for strong Ohmic resistivity at least second-order convergence.  For $\eta_{\rm OR} \leq 0.001$, the errors do not decrease for low resolution which can be explained by the fact that the numerical resistivity is dominating and therefore the profile in the simulation corresponds to one with larger $\eta_{\rm OR}$. By further increasing the resolution we can find for $\eta_{\rm OR} = 0.001$ decreasing errors, although we cannot further increase the resolution since otherwise the simulations become unstable.

\subsection{Artificial resistivity}
\label{subsec:artResivisity}

Based on a better understanding of the stability of our implementation for the Hall effect we opt for an approach where we deliberately increase the magnetic diffusivity by the small amount required for stabilization. This prevents us from using the HLL-modified Riemann solver in the full domain. Instead we locally increase the Ohmic resistivity. The condition for stability of the Hall effect can be formulated as
\begin{equation}
   \left| \eta_{\rm H} \right|<  C_1\left(\eta_{\rm OR} +  \eta_{\rm A} +  \eta_{\rm N}\right),
   \label{eq:stabilityHall1}
\end{equation}
where we introduced a coefficient $C_1$ of order unity. The contribution of the spatial discretization to the numerical resistivity can in principle be parameterized by \citep{Rembiasz2017}:
\begin{equation}
     \eta_{\rm N,i} = C_2 \left(\frac{\Delta x_i}{L}\right)^2 c_w L,
\end{equation}
for a second order accurate scheme, where $C_2$ is another parameter of order unity, $\Delta x_i =  \left[ 3V_i / \left(4 \pi\right) \right]^{1/3}$ is the approximate size of a cell $i$, $L$ denotes the characteristic length scale of the problem, and $c_w$ gives the local wave speed, which we set equal to the velocity of the fast magneto-sonic wave, $c_w = \sqrt{c_s^2+B^2  / \rho}$. But as we have seen in the last section, the factor $C_2$ can also depend on $\Delta x$ in our scheme, and in general it is not always obvious what the characteristic length scale of a problem is \citep{Rembiasz2017}.

This makes its hard for us to find the required lower boundary for $C_2 L$.  We will therefore neglect the numerical diffusivity in equation~(\ref{eq:stabilityHall1})  and instead use the condition:
\begin{equation}
   \left| \eta_{\rm H} \right|<  C_1\left(\eta_{\rm OR} +  \eta_{\rm A} \right).
   \label{eq:stabilityHall2}
\end{equation}
Following the results presented in the last section we use $C_1 = 5$, which is also found to be stable for all simulations of the collapse of a protostellar core we present later in the paper. We note that by neglecting the numerical diffusivity we also apply some artificial resistivity in regions where $\eta_N$ is already sufficient to stabilize the Hall term, but in this case we also expect $\eta_{\rm OR} < \eta_N$, and therefore this should not influence our results.

\section{Propagation of a whistler wave}
\label{subsec:PropagationWhistler}

As we have seen in Section~\ref{subsec:linearWaveMHDHall}, the Hall effect introduces two additional MHD waves. We perform here a similar study as in \cite{marchand2018impact}, i.e.~we try to measure the numerical dispersion relation and the damping rates for those waves. We will simulate a polarized, travelling wave in the $x$-direction with a guide field $B_x$ and wave number $k$. By assuming $B_{y,z} \ll B_x$ the Hall coeffcient $\eta_{\rm H}$ becomes constant and the Hall effect leads to a modulation of the amplitude of the magnetic field with frequency:
\begin{equation}
    \sigma = \frac{\eta_{\rm H}}{2} k^2.
\end{equation}
The frequency of the oscillation is modified to:
\begin{equation}
    \omega = \sqrt{v_A^2 k^2 + \sigma^2}.
\end{equation}
Both frequencies are linear combinations of the two frequencies defined in (\ref{eq:dispersionRelationWhistler}), and therefore a Fourier decomposition of the temporal signal allows use to measure the dispersion relation for the whistler and the ion cyclotron waves. We simulate a cubic box of size $L_x \times L_y \times L_z = 1 \times 10 /N \times 10N$ with uniform density and pressure. $N^3$ is here again the number of cells we expect in a cubic box with size unity.
The initial conditions are given by:
\begin{align}
    \rho =1,\:
    P = 1,\:
    B_x = 0.1 /\sqrt{4\pi},\:
    B_y = 10^{-3}/\sqrt{4\pi} \cos\left( k x\right), \nonumber\\
    B_z = 0, \:
    v_x = 0, \:
    v_y = \delta v \cos\left(k x\right),\:
    v_z = \delta v \frac{\sigma}{\omega} \sin\left(k x\right),
\end{align}
and we use periodic boundary conditions, which restricts $k$ to be an integer times $2\pi$. We use an adiabatic equation of state with $\gamma = 5/3$, and the full analytical solution including $\delta v$ is given in Appendix~\ref{app:AlvefnWave}.

\subsection{Linear dispersion relation}
\label{subsubsec:linearDispersionRelation}

\begin{figure}
    \centering
    \includegraphics[width=1\linewidth]{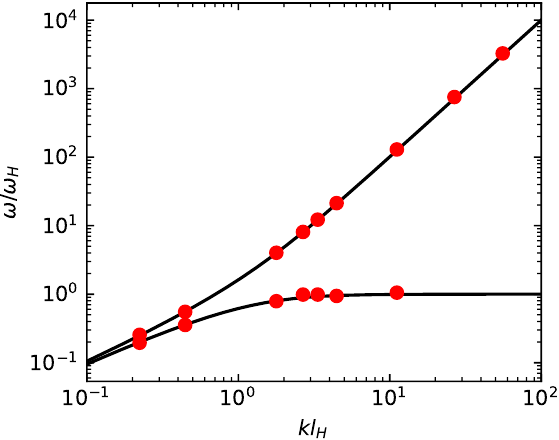}
    \caption{The dispersion relation of the whistler and ion cyclotron waves (black line, see equation \ref{eq:dispersionRelationModified}) and the measured relation in our simulations (red symbols). More information can be found in Section~\ref{subsubsec:linearDispersionRelation}.}
    \label{fig:whistler_dispersion}
\end{figure}

As a first test we try to measure the dispersion relation (\ref{eq:dispersionRelationWhistler}) using a resolution of $160\times 10 \times 10$ cells initially ($N=160$). We introduce the Hall frequency $\omega_H = v_A^2 / \eta_{\rm H}$, which allows us to rewrite equation (\ref{eq:dispersionRelationWhistler}) as:
\begin{equation}
    \frac{\omega}{\omega_H} = \pm \frac{\left(k l_H\right)^2}{2} +\sqrt{\left(\frac{k^2 l_H^2}{2}\right)^2 + \left(k l_H\right)^2}.
    \label{eq:dispersionRelationModified}
\end{equation}
The normalized frequency is now only a function of the product $kl_H$, with $l_H = \eta_{\rm H} / v_A$. This product is the inverse of the Lundquist number we introduced in equation (\ref{eq:lundquistNumber}). In \cref{tab:OverViewDispersionRelation} we give an overview of the parameters we used in the simulations to measure the dispersion relation and  \cref{fig:whistler_dispersion} shows our results. We find perfect agreement with the theoretical dispersion relation. For large $k l_H$ we are only able to measure the high-frequency branch since the lower frequencies require a much longer measurement.

 \begin{table}
    \centering
    \begin{tabular}{c|c|c|c|c}
    \hline
        $k/\left(2 \pi\right)$ & $\eta_{\rm H}$ & $k l_H$ & $T_{\rm max}$ & $dt$ \\
        \hline
2 & 0.0005 & 0.22 & 400 &0.0500\\
2 & 0.001 & 0.45 & 200 & 0.0250\\
4 & 0.002 & 1.78 &100& 0.0125\\
6 & 0.002 & 2.67 &80& 0.0100\\
3 & 0.005 & 3.34 &200& 0.0250\\
4 & 0.005 & 4.45 &100&0.0125 \\
5 & 0.01 & 11.14 &150 & 0.0200\\
6 & 0.02 & 26.73 & 25 & 0.0025\\
5 & 0.05 & 55.68& 50 & 0.0050 \\
     \hline

    \end{tabular}
    \caption{Overview of simulations we performed to measure the dispersion relation of the Hall MHD waves.
    We give the wave vector $k$, the Hall diffusivitiy $\eta_{\rm H}$, the product $kl_H$, the total simulation time $T_{\rm max}$ and the minimum resolved time $dt$ to measure the frequencies.}
    \label{tab:OverViewDispersionRelation}
\end{table}

\subsection{Diffusivity of our scheme}
\label{subsec:waveDiffusivity}

\begin{figure}
    \centering
    \includegraphics[width=1\linewidth]{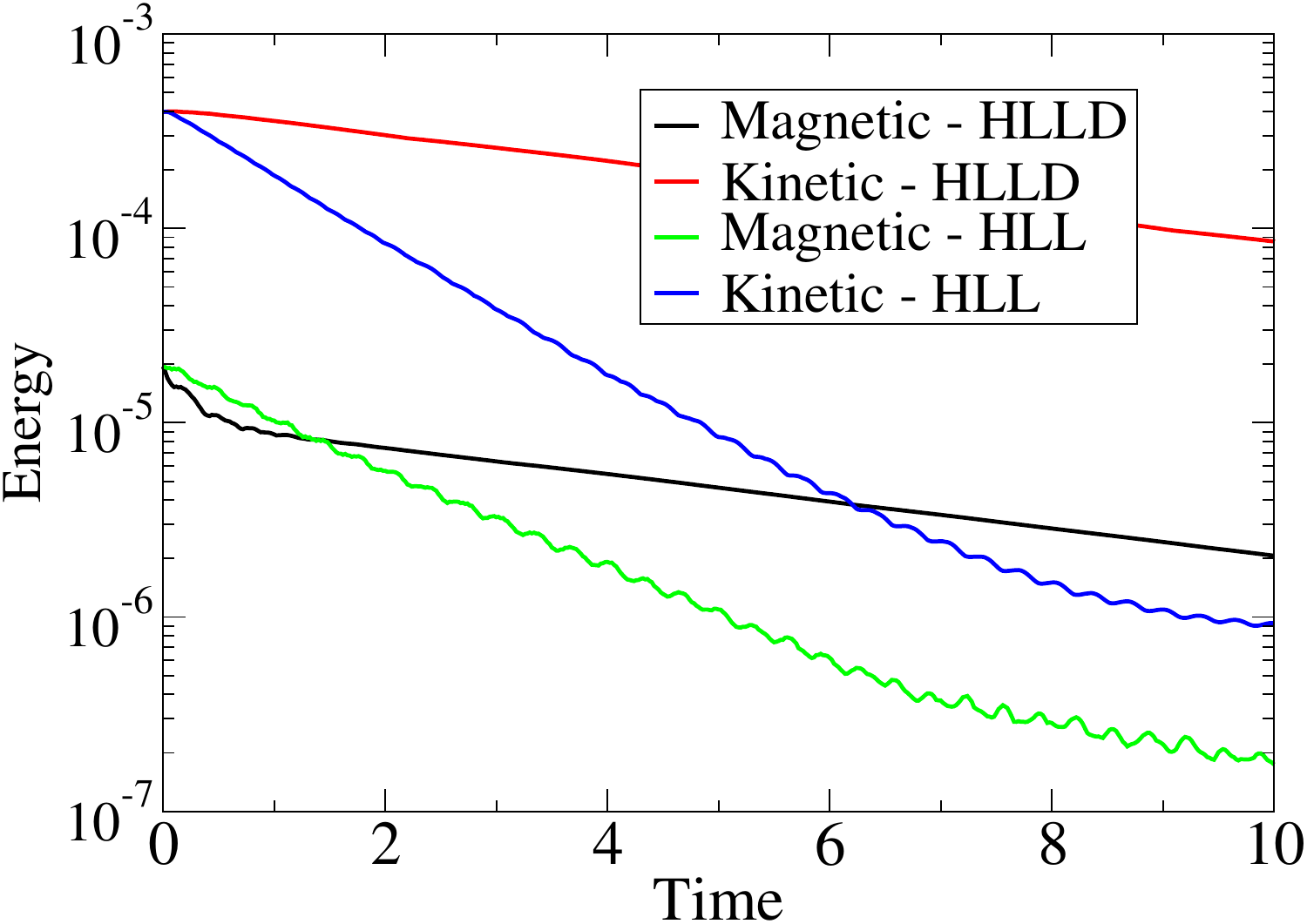}
    \caption{The temporal evolution of the magnetic and kinetic energy for simulations of an MHD wave with $k=10 \left(2\pi\right)$ and resolution $N=160$. We perform the simulation once with the HLLD Riemann solver together with Ohmic diffusion and once with the HLL Riemann solver.}
    \label{fig:EnergyEvolutionK10LR}
\end{figure}

Both explicit and numerical diffusion lead to a damping of the amplitude of the wave. The damping rate is in general a function of the resolution and the wave vector $k$, which we will further analyze in this section. For comparison, we also implemented the Hall modified HLL solver from \cite{lesur2014thanatology}, which we also shortly discuss in Appendix~\ref{app:HallModifiedHLL}. As an example, we show in \cref{fig:EnergyEvolutionK10LR} the temporal evolution of the magnetic and kinetic energy for a setup with $N= 160$, $k=10 \left(2\pi\right)$ and $\eta_{\rm H} = 0.005$. At the beginning, we can observe for the HLLD Riemann solver a strong decay of the magnetic field, and at a later time a decay of the magnetic field with the same decay rate as the kinetic energy. For the HLL Riemann solver both types of energy decay at a similar rate but in general significantly faster than for the HLLD Riemann solver.  The faster decay of the kinetic energy in this case can also be explained by the more diffusive behaviour of the HLL Riemann solver since it also increases the numerical viscosity.

\begin{figure}
    \centering
    \includegraphics[width=1\linewidth]{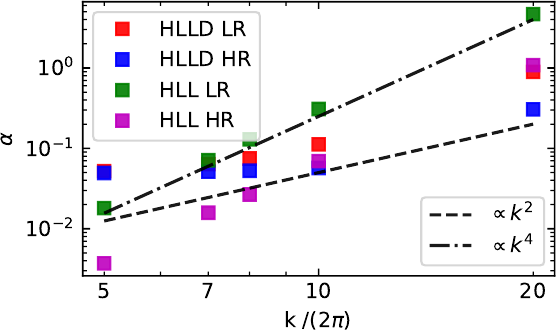}
    \caption{The damping rate $\alpha$ of the magnetic field as a function of the wave vector $k$. We performed simulations with $N=160$ (LR) and $N=320$ (HR), as well as with HLLD and HLL Riemann solver.}
    \label{fig:damping_factor_k_function}
\end{figure}

The evolution of the energy can be approximated by an exponential decay with a damping rate $ \alpha$, which we define as half of the decay rate of the magnetic energy.
By fitting a function proportional to $e^{- 2 \alpha t}$ to the evolution of the magnetic field energy, we can measure this decay rate as a function of $k$ and resolution.
We perform simulations with $N= 160$ and $N=320$, different $k$ and $\eta_{\rm H} =0.005$, and show the results in \cref{fig:damping_factor_k_function}.
As expected for the HLL Riemann solver the damping rate is proportional to $k^4$ \citep{marchand2018impact} and decreases by a factor of $4$ if we double the resolution.
For the HLLD Riemann solver, the damping rate is a significantly weaker function of $k$.
For small $k$ it is approximately independent of the resolution since in this case the damping is dominated by the Ohmic diffusion. 
For larger $k$ the numerical diffusion plays a role and the damping rate can be reduced by increasing the resolution.

\begin{figure*}
    \centering
    \includegraphics[width=1\linewidth]{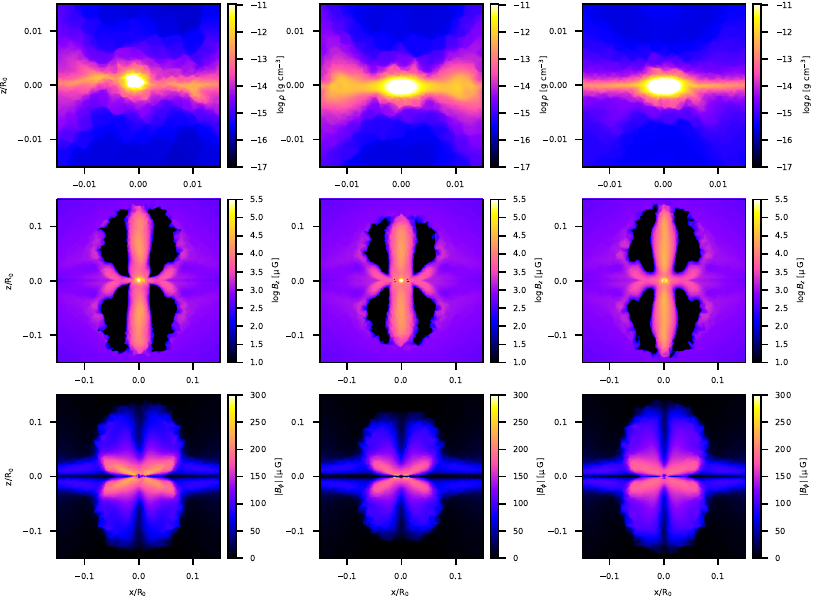}
    \caption{Formation of a dense core in the collapse of a magnetized cloud under its own self-gravity with different models for the magnetic field. The left column shows a simulation with ideal MHD, while the middle one adds a constant positive, and the right one a constant negative Hall coefficient as well as Ohmic diffusion for stability. All the panels show projected slices in the $xz$-plane with the slice having a thickness of 0.1 times the side length of the plot. In the top row, we display a volume-weighted density projection of the gas in the central region (0.03 $R_0$ in diameter). The middle and bottom row show density-weighted projections of the $z$- and the azimuthal component of the magnetic field, respectively, in a region 10 times as large. Note that the $z$-component of the magnetic field is negative (inverted with respect to the initial condition) due to the outflow in some regions. Results from the three simulations are shown after an identical amount of elapsed time, approximately $3.4 \times 10^4$  years.
    }
    \label{fig:Collapse}
\end{figure*}

\section{Collapse of a magnetized cloud}
\label{subsec:Collapse}

To test our new module for non-ideal MHD in an environment closer to  typical science applications, we repeat in this section the simulation of the collapse of a rotating, magnetized cloud under its own gravity performed in Paper~I but now with the Hall effect. This setup acts as a simplified model for star formation, allowing us to test the stability of our methods in a situation with a large dynamic range of scales. 

The initial conditions are again taken from \cite{pakmor2011magnetohydrodynamics} and \cite{marinacci2018non}, which were in turn adapted from \cite{hennebelle2008magnetic}. They consist of a rigidly rotating, homogeneous sphere with radius $R_0 = 0.015\,\mathrm{pc}$ and mass $1\,{\rm M}_\odot$ embedded via a small transition region in a low density environment with a density contrast of $100$. The total size of the simulation box is $0.06\,\mathrm{pc}$. The initial density of the cloud is $4.8 \times 10^{-18} \mathrm{g\ cm^{-3}}$, which translates to a free fall time of $3 \times 10^4$ years. We choose a ratio of $0.045$ between rotational and gravitational energy, which is equivalent to a rotational period of $4.7 \times 10^5$ years for the rigid body rotation around the $z$-axis. The cloud is initially penetrated by a homogeneous, purely vertical magnetic field of magnitude $107\,\mathrm{\mu G}$, equivalent to a mass-to-flux over critical mass-to-flux ratio of $5.6$. We use a barotropic equation of state \citep{hennebelle2008magnetic}
\begin{equation}
    P = \rho c_{s,0}^2 \sqrt{1 + \left(\rho / \rho_c \right)^{4/3}},
\end{equation}
with sound speed $c_{s,0} = 0.2\, \mathrm{km\ s^{-1}}$ and critical density $\rho_c = 10^{-13}\,\mathrm{g\ cm^{-3}}$, which represents a transition from an initially isothermal EOS to an adiabatic one at higher densities. We start with an initial resolution of $128^3$ cells on a uniform Cartesian mesh and apply periodic boundary conditions at the edges of the simulation box for all quantities except gravity. Cells are refined if their free fall time is smaller than ten times their sound crossing time. We also introduce a minimum volume of $5 \times 10^{-17} \mathrm{pc}$ which is equivalent to an effective resolution of $16384^3$ cells.  The latter condition reduces the computational cost, especially considering the quadratic dependence of the diffusive timestep on cell radius. Here, we set a constant coefficient of $\eta_{\rm H} = \pm 10^{18} \mathrm{cm^2\ s^{-1}}$ to demonstrate the aforementioned bimodality of the Hall effect in either promoting or suppressing disk formation. To stabilize our simulation we also use $\eta_{\rm OR} = 2 \cdot 10^{17} \mathrm{cm^2\ s^{-1}}$.

The drift of magnetic field lines in comparison to ideal MHD due to the Hall effect is 
\begin{equation}
v_{\rm drift} = - \frac{\eta_{\rm H} \, {\bm J}}{\vert \bm B \vert}.
\end{equation}
In our setup, the initial rotation and magnetic field axes are aligned. In this case, the azimuthal current resulting from the radial bending of the magnetic field is parallel to the rotational velocity of the gas. Depending on the sign of $\eta_{\rm H}$, the magnetic braking caused by the azimuthal bending of the magnetic field is either weakened ($\eta_{\rm H} > 0$) or further enhanced ($\eta_{\rm H} < 0$) by the presence of the Hall effect. The effect of this bimodality is shown in Figure \ref{fig:Collapse}, where positive and negative Hall diffusivity show diverging trends with respect to ideal MHD. In the case with positive coefficient, the drift of magnetic field lines caused by the Hall effect leads to a weaker azimuthal field and thereby to a broader and slower outflow compared to ideal MHD. On the other hand, the azimuthal field is enhanced and the outflow is both faster and more strongly collimated if the coefficient is negative. The effect on the azimuthal field can in Fig.~\ref{fig:Collapse} most clearly be seen in the fact that in the bottom panels a much larger section around the midplane has a very small value (black in the image) in the simulation with the positive coefficient than with the nagative one.

Note that when the Hall coefficient is (locally) given by a chemical model, the bimodality is instead usually characterized by an (anti-)alignment of the axes of rotation and magnetic field. For instance, \cite{wurster2021impact} show that since in their model the Hall-coefficient is negative throughout the early evolution of envelope and disk, an aligned configuration of magnetic field and rotation axes enhances the azimuthal field and removes angular momentum from the disk -- this corresponds to the situation in the right panel of Figure \ref{fig:Collapse}. They even find a counter-rotating inner disk at the end of the simulation. The middle panels can on the other hand be seen as a simplified model of the anti-aligned configuration (in our setup the signs of the Hall coefficient and the current are both reversed with respect to theirs, cancelling out). Similarly, \cite{zhao2020hall} observe either enhanced or suppressed disk formation, depending on alignment. \cite{Bai2017}, in contrast to the two previously mentioned works, assume a positive Hall coefficient throughout their disks and consequently find instead enhanced outward angular momentum transport in the anti-aligned configuration.

\section{Summary and Conclusions}
\label{sec:discussionSummary}

In this paper we have extended the non-ideal MHD module in the moving mesh code {\small AREPO} to include the Hall effect. The Hall effect is not diffusive but dispersive, leading to the introduction of two new MHD waves, the whistler and the ion-cylcotron waves. It is known to destabilise second order accurate time integration schemes without sufficient magnetic diffusion, leading to various approaches in the past to reach stability by increasing the diffusion.

Our implementation of the Hall effect follows the method we introduced in \cite{zier2023} for modelling Ohmic and ambipolar diffusion, which relies on a least-squares fit to estimate the magnetic field gradient at the interfaces between neighbouring cells. By simulating a magnetic C-shock, we show in Section~\ref{sec:stabilityHall} that our scheme suffers from the typical numerical instability at high resolution. By adding sufficient Ohmic resistivity we can stabilize our method and obtain second-order accuracy. The critical ratio below which we find the instability for our scheme is $\eta_{\rm OR} / \eta_{\rm H} < 0.1$. To be on the safe side, we therefore conservatively propose to enforce the condition $\eta_{\rm OR} \geq \eta_{\rm H} / 5$ in all cells. For this choice and for these parameters, we did not find any signs of instability. 

Using this new condition, we have shown in Section~\ref{subsec:PropagationWhistler} that we are able to accurately resolve the dispersion relation of the new MHD waves that appear for the Hall effect. We compared the damping rate for the magnetic field with a similar implementation using a modified HLL Riemann solver instead of an artificially increased Ohmic diffusion. We showed that the increased numerical diffusion of the HLL Riemann solver is particularly efficient on small scales, but also increases the numerical viscosity. As a first example of a more complicated setup, we simulated in Section~\ref{subsec:Collapse} the collapse of a magnetised gas cloud with a constant Hall diffusion coefficient. We demonstrated that, depending on the coefficient of the Hall effect, the magnetic braking can be weakened or strengthened.

We developed our module with the collapse of protostellar cores in mind, since in this case we can directly profit from the high accuracy and adaptive resolution of the moving mesh approach. In these simulations, the resolution within a potentially forming disk is rather low, and we need a low numerical viscosity to accurately follow their evolution. The Hall effect may only dominate within small regions of the system \citep{wurster2018role}, and therefore we feel that using the more diffusive HLL Riemann solver globally is excessive, rather we consider a localized increase of the Ohmic diffusion preferable. We are currently also working on more realistic simulations with non-constant diffusivities  (Mayer et al., in preparation). For the future we also plan to implement a third-order accurate time integration scheme for the Hall effect similar to the one we already implemented in \cite{zier2022simulating} for ideal MHD. This scheme might be stable without requiring any artificially increased diffusion.

\section*{Acknowledgements}

The authors acknowledge helpful discussions with R\"udiger Pakmor.

\section*{Data Availability}
The data underlying this paper will be shared upon reasonable request to the corresponding author.

\begin{appendix}

\renewcommand{\thefigure}{A\arabic{figure}}
\setcounter{figure}{0}

\bibliographystyle{mnras}
\bibliography{main.bib}

\section{Analytical solutions of test problems}
\label{app:analyticalSolutions}

\subsection{C-shock}
\label{app:CShock}
We mostly follow the method described in \cite{marchand2018impact} to construct the equilibrium state of a C-shock in the $x$-direction.
We start with equation (\ref{eq:nonidealMHDEquations}) and remove all derivatives except $\frac{\partial}{\partial x}$:
\begin{gather}
    \frac{\partial \rho v_{x}}{\partial x} = 0,\\
    \frac{\partial \left(\rho v_x^2 + \rho c_s^2 +B^2/2\right)}{\partial x} = 0,\\
     \frac{\partial \left(\rho v_x v_y -B_x B_y\right)}{\partial x} = 0,\\
     \frac{\partial \left(\rho v_x v_z -B_x B_z\right)}{\partial x} = 0,\\
     \frac{\partial B_x}{\partial x} = 0, \label{eq:BxCshock}\\
     \frac{\partial}{\partial x}\begin{pmatrix}
     v_x B_y - v_y B_x\\
     v_x B_z - v_z B_x
     \end{pmatrix}
     \label{eq:inductionEquationCShock}
     =\\
     \frac{\partial}{\partial x}\begin{pmatrix}
     \left[\eta_{\rm OR} +\eta_{\rm AD} \left(1 - \frac{B_z^2}{B^2}\right)\right] \frac{\partial B_y}{\partial x} + 
     \left[\eta_{\rm H}\frac{B_x}{B} +\eta_{\rm AD}\frac{B_y B_z}{B^2}\right] \frac{\partial B_z}{\partial x} \nonumber\\
    \left[-\eta_{\rm H}\frac{B_x}{B}+\eta_{\rm AD}\frac{B_y B_z}{B^2}\right] \frac{\partial B_y}{\partial x} + 
     \left[\eta_{\rm OR} +\eta_{\rm AD} \left(1 - \frac{B_y^2}{B^2}\right)\right] \frac{\partial B_z}{\partial x}
     \end{pmatrix} .
\end{gather}
From equation (\ref{eq:BxCshock}) it follows that $B_x$ is constant, and we can thus solve equation (\ref{eq:inductionEquationCShock}) by rewriting it as a matrix equation and performing a matrix inversion:
\begin{gather}
    \frac{\partial B_y}{\partial x} = \frac{M_{1}R_{22}-M_2 R_{12}}{R_{11} R_{22} -R_{21} R_{12}},\\
     \frac{\partial B_z}{\partial x} = \frac{M_{2}R_{11}-M_1 R_{21}}{R_{11} R_{22} -R_{21} R_{12}},
\end{gather}
 where we introduced the vector
 \begin{equation}
     M = \begin{pmatrix}
     v_x B_y - v_y B_x + v_{y,0} B_{x,0} - v_{x,0} B_{y,0} +M_{1,0}\\
     v_x B_z - v_z B_x + v_{z,0} B_{x,0} - v_{x,0} B_{z,0} +M_{2,0}
     \end{pmatrix}
 \end{equation}
 and the matrix
 \begin{equation}
     R =  \begin{pmatrix}
     \eta_{\rm OR} +\eta_{\rm AD} \left(1 - \frac{B_z^2}{B^2}\right) & \eta_{\rm H}\frac{B_x}{B} +\eta_{\rm AD}\frac{B_y B_z}{B^2}\\
     -\eta_{\rm H}\frac{B_x}{B}+\eta_{\rm AD}\frac{B_y B_z}{B^2} & \eta_{\rm OR} +\eta_{\rm AD} \left(1 - \frac{B_y^2}{B^2}\right)
     \end{pmatrix}.
 \end{equation}
 Quantities with subscript $0$ denote the initial state, and the vector
 \begin{equation}
     M_0 = \begin{pmatrix}M_{1,0}\\M_{2,0} \end{pmatrix} = \begin{pmatrix}R_{11}\left.\frac{\partial B_y}{\partial x}\right|_{x=0} + R_{12} \left.\frac{\partial B_z}{\partial x}\right|_{x=0} \\R_{21}\left.\frac{\partial B_y}{\partial x}\right|_{x=0} + R_{22} \left.\frac{\partial B_z}{\partial x}\right|_{x=0} \end{pmatrix}
 \end{equation}
 is required to trigger the shock.
 The conserved quantities
 $Q= \rho v_x $
 and $K = Q v_x + \frac{Q}{v_x}c_s^2 + \frac{B^2}{2}$ allow us to calculate the remaining unknowns after we derive the magnetic field at the new position:
 \begin{gather}
     v_x = \frac{1}{2Q} \left[K - \frac{B^2}{2} - \sqrt{\left(K-\frac{B^2}{2}\right)^2 - 4 Q^2 c_s^2} \right]\\
     v_y = v_{y,0}+ \frac{B_x \left(B_y - B_{y,0}\right)}{Q}\\
     v_z = v_{z,0}+ \frac{B_x \left(B_z - B_{z,0}\right)}{Q}\\
     \rho = \frac{Q}{v_x}.
 \end{gather}
 In practice, we use a fourth order Runge Kutta integrator and start at $x= x_0$ with the post shock values and integrate backwards until we reach $x=0$. We first calculate the magnetic field at a new position and then update the other quantities. For easier reproducibility we made the code publicly available\footnote{\url{https://github.com/ChessOli/C-shock-Structure}}.
 
 The equations for the static state of the C-shock are invariant under the transformation:
 \begin{equation}
     \begin{pmatrix}
     \eta_{\rm H}\\
     \eta_{\rm OR}\\
     \eta_{\rm AD}\\x
     \end{pmatrix}
     \longrightarrow 
     C      \begin{pmatrix}
     \eta_{\rm H}\\
     \eta_{\rm OR}\\
     \eta_{\rm AD}\\x
     \end{pmatrix}
 \end{equation}
 for a constant $C$. This means that in the simulations with $\eta_{\rm AD} = 0$  discussed in this paper the structure of the C-shock is determined by the ratio $ \eta_{\rm OR}/\eta_{\rm H}$.
  
  \subsection{Modulated Alfvén wave}
  \label{app:AlvefnWave}
  For the setup discussed in Section~\ref{subsec:PropagationWhistler} we find the analytical evolution:
\begin{eqnarray}
    B_y &=& \delta B \cos\left(\omega t - k x\right) \cos\left(\sigma t\right) \\
    B_z &=& \delta B \cos\left(\omega t -k x\right) \sin\left(\sigma t\right) \\
    v_y & = & \delta v \left[\cos\left(\sigma t\right) \cos\left(\omega t - k x\right) + \frac{\sigma}{\omega} \sin\left(\sigma t\right) \sin\left(\omega t - k x\right) \right]\\
    v_z & = & \delta v \left[\sin\left(\sigma t\right) \cos\left(\omega t-  k x\right) - \frac{\sigma}{\omega} \cos\left(\sigma t\right) \sin\left(\omega t - k x\right) \right] 
\end{eqnarray}
with an amplitude of the velocity perturbation equal to
\begin{equation}
    \delta v = \frac{k}{\rho_0} B_0 \delta B \frac{-\omega}{\omega^2 - \sigma^2} .
\end{equation}

 \section{Influence of numerical details on the collapse of protostellar cores}
 \label{app:RiemannColappse}

\subsection{The Hall modified HLL Riemann solver in AREPO}
\label{app:HallModifiedHLL}

To overcome the inherent stability problems of the operator-splitting approach, we have alternatively implemented a modified HLL Riemann solver to incorporate the Hall effect directly into the ideal MHD flux calculation. \cite{toth2008} introduced this method which was later implemented in \cite{lesur2014thanatology} in the {\small PLUTO}  code.
We follow their implementation and solve the equation:
\begin{equation}
    \frac{\partial \bm U }{\partial t}+ \nabla \cdot \left[\bm F_{\rm ideal}(\bm U) +  \bm F_{\rm H}(\bm U)\right]   =0
    \label{eq:HLLEquation},
\end{equation}
which is problematic because the flux depends on the magnetic field and its derivative via the current $\bm J = \nabla \times \bm B$. This makes the Riemann problem ill-defined, since $\bm J$ is not defined if there is a discontinuity in the magnetic field.  To get around this problem, we treat $\bm J$ as an external parameter to the Riemann problem and estimate it as the average of the currents of the two cells of the interface, using the standard gradient estimates in {\small AREPO}.   

As other studies in the past \citep{lesur2014thanatology,Bai2017,marchand2018impact,marchand2019} we have not been able to modify the accurate HLLD or Roe Riemann solvers to deal with this ill-defined Riemann problem and therefore we use the diffusive HLL Riemann solver as a basis instead. The HLL Riemann solver requires only the smallest algebraic signal velocity of the left state and the largest algebraic signal velocity of the right state. We define them as:
\begin{equation}
S_{\rm L,R} \equiv v \pm {\rm max}(c_{\rm f},c_{\rm w}),
\end{equation}
where $v$ is the fluid velocity, $c_{\rm f}$ is the fast magnetosonic speed and $c_{\rm w}$ is the whistler wave speed. Since the Hall effect is dispersive, we use the cell radius $r$ to calculate the fastest whistler wave.

For the HLL Riemann solver the flux to update the conservative variables is given by:
\begin{displaymath}
{\bm F}^*= {\bm F}_{\rm L} \quad  {\rm if} \, S_{\rm L} > 0;
\end{displaymath}
\begin{displaymath}
{\bm F}^*= {\bm F}_{\rm R} \quad {\rm if} \, S_{\rm R} < 0;
\end{displaymath}
\begin{displaymath}
{\bm F}^*= \frac{S_{\rm R} S_{\rm L} ({\bm U}_{\rm R} - {\bm U}_{\rm L}) + S_{\rm R} {\bm F}_{\rm L} - S_{\rm L}{\bm F}_{\rm R}}{S_{\rm R} - S_{\rm L}} \quad {\rm else},
\end{displaymath}
where $F_{\rm L,R}$ are the left and right fluxes. In contrast to our standard scheme with the HLLD Riemann solver and Ohmic diffusion, this method reduces the accuracy even in the absence of the Hall effect.

\subsection{Comparison of the HLL and HLLD Riemann solver}

\begin{figure}
    \centering
    \includegraphics[width=1\linewidth]{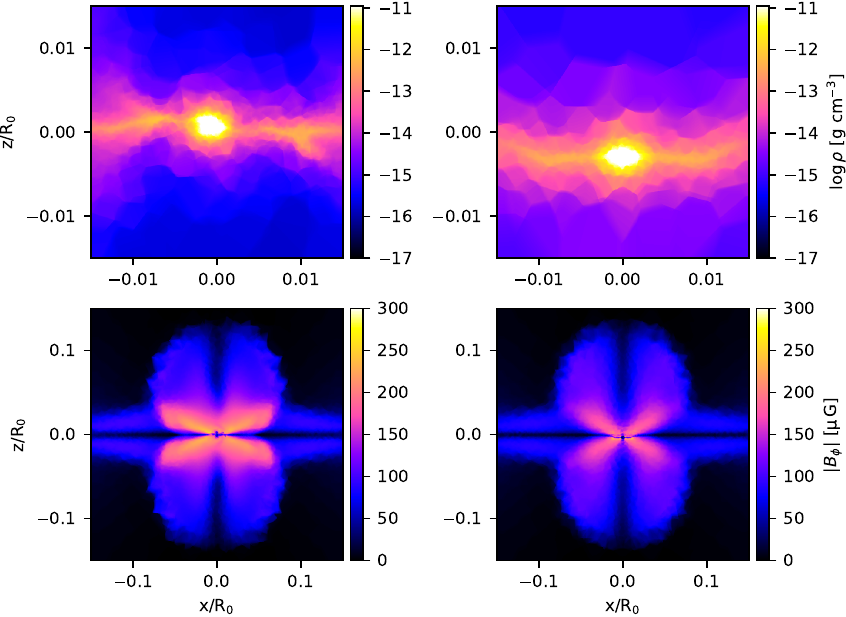}
    \caption{A comparison of the density (top) and azmithuthal magnetic field (bottom) structure in ideal MHD simulations of the collapse of a magnetized cloud core with the HLLD Riemann solver (left side) and the HLL Riemann solver (right side).
    We use the same parameters as we use in \cref{fig:Collapse} to produce the plot.
    }
    \label{fig:Collapse_hll_hlld}
\end{figure}

\begin{figure}
    \centering
    \includegraphics[width=1\linewidth]{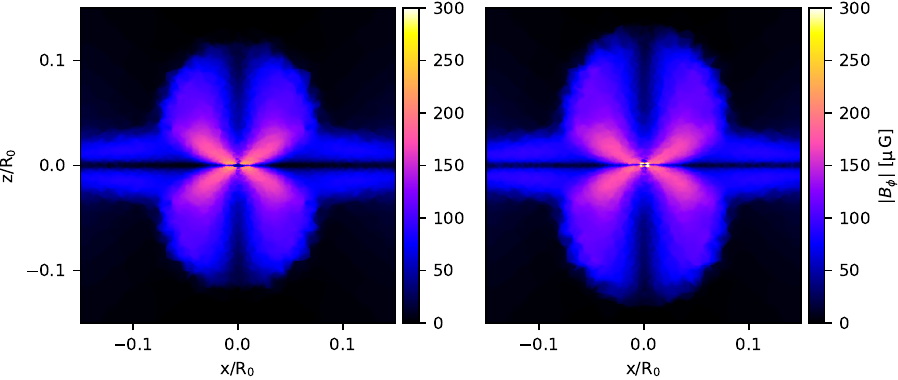}
    \caption{Slices of the azimuthal magnetic field in simulations with the same setup as in \cref{fig:Collapse} but with the Hall modified HLL Riemann solver instead of the HLLD Riemann solver with additional Ohmic diffusion.
    We use on the left side $\eta_{\rm H} = 10^{18} \mathrm{cm^2\ s^{-1}}$ and on the right side $\eta_{\rm H} = - 10^{18} \mathrm{cm^2\ s^{-1}}$.
    The elapsed time and parameters of the plot are identical to the bottom panels of \cref{fig:Collapse}. 
    }
    \label{fig:HLL_Collapse}
\end{figure}

To better understand the influence of the Riemann solver on our results we repeated the simulations presented in Section~\ref{subsec:Collapse} with the Hall modified HLL Riemann solver instead of an artificial Ohmic diffusion. As we can see in \cref{fig:Collapse_hll_hlld} the HLL Riemann solver not only is more diffusive for the magnetic field but also directly influences hydrodynamic quantities such as the density. We also repeated the simulations with the Hall effect and show in  \cref{fig:HLL_Collapse} the azimuthal magnetic field. In comparison to the simulations with HLLD Riemann solver, the influence of the Hall effect is also reduced, showing only very small differences in the outflow structure. While the general trends expected from the inclusion of the Hall effect are still captured by the modified HLL solver, it appears that the increased diffusivity relative to the HLLD solver that we showed in \cref{fig:EnergyEvolutionK10LR} for the wave test also has a significant effect in the context of more complicated applications. 

\subsection{Effect of the artificially increased Ohmic diffusion}

\begin{figure}
    \centering
    \includegraphics[width=1\linewidth]{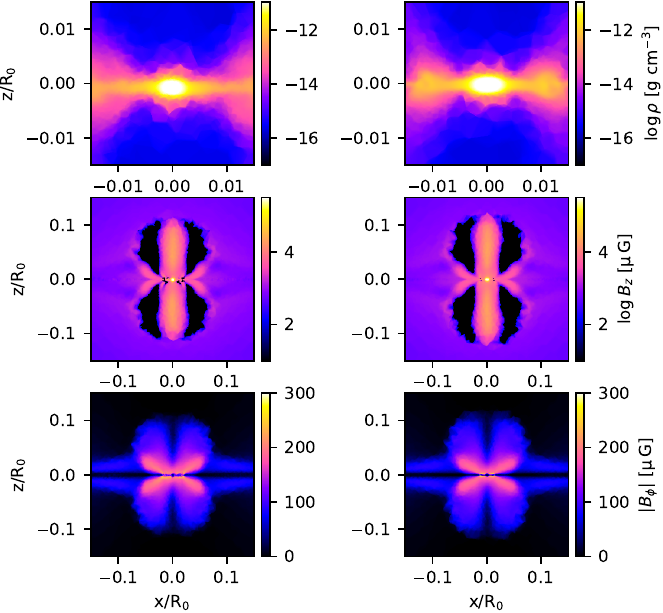}
    \caption{Density and magnetic field slices in simulations of the collapse of a magnetized cloud using different resistivity coefficients. The left panels have solely $\eta_{\rm H} = 10^{18}\,\mathrm{cm^2\ s^{-1}}$, while the right panels show simulations that additionally include $\eta_{\rm OR} = 2 \times 10^{17}\,\mathrm{cm^2\ s^{-1}}$.
    }
    \label{fig:Collapse_with_ohm_1e18}
\end{figure}

As we have discussed in Section~\ref{sec:stabilityHall}, simulations with the Hall effect become unstable if there is too little diffusion (numerical or physical) to stabilize it.  But by adding an unphysical Ohmic diffusion we also might influence the outcome of our simulations. We therefore performed simulations with $\eta_{\rm H} = 10^{18}\mathrm{cm^2\ s^{-1}}$ for which the numerical resistivity of the HLLD Riemann solver alone is high enough to stabilize the code for our resolution. Nevertheless, as one can see in \cref{fig:Collapse_with_ohm_1e18} there is some noise observable in the magnetic field. We repeated the simulation with an additional Ohmic diffusion $\eta_{\rm OR} =2\times 10^{17}\mathrm{cm^2\ s^{-1}}$ and show in \cref{fig:Collapse_with_ohm_1e18} that this additional diffusion removes the noise and leads to a smoother magnetic field. But the large scale structure is still very similar. The additional Ohmic diffusion reduces the efficiency of magnetic braking which leads to a stronger disk like structure observable in the density field.

We note that in simulations with realistic values for the diffusivity coefficients, the Hall effect is dominated almost everywhere by either ambipolar diffusion (low density) or Ohmic diffusion (high density). Therefore, instabilities resulting from the Hall effect are suppressed in most situations by the presence of strong physical diffusion. Otherwise, they can be mitigated by locally increasing the diffusion above its physical value, ensuring that the threshold for stability ($\eta_{ \rm OR , AD} > 0.2 \, \vert \eta_{\rm H} \vert$ based on our C-shock tests) is guaranteed everywhere.

\end{appendix}
\bsp	
\label{lastpage}
\end{document}